\magnification = \magstep1
\baselineskip = 24 true pt
\hsize = 16 true cm
\vsize = 22 true cm
\def\dddot#1{{\mathop{#1}\limits^{\vbox to-1.4ex{\kern-2ex%
   \hbox{...}\vss}}}}
\centerline{\bf THE MAXIMAL KINEMATICAL INVARIANCE GROUP OF FLUID} 
\centerline {\bf DYNAMICS AND EXPLOSION-IMPLOSION DUALITY}
\medskip
\medskip
\centerline {L. O'Raifeartaigh\footnote{${}^{(a)}$}{\it Deceased} and V. V. 
Sreedhar\footnote{${}^{(b)}$}{\it sreedhar@stp.dias.ie}}
\centerline {School of Theoretical Physics}
\centerline {Dublin Institute for Advanced Studies}
\centerline {10, Burlington Road}
\centerline {Dublin 4, Ireland}
\medskip
\centerline {\bf Abstract}

It has recently been found that supernova explosions can be simulated in the 
laboratory by implosions induced in a plasma by intense lasers. A theoretical 
explanation is that the inversion transformation, ($\Sigma: t \rightarrow -1/t,~
{\bf x}\rightarrow {\bf x}/t$), leaves the Euler equations of fluid dynamics, 
with standard polytropic exponent, invariant. This implies that the kinematical
invariance group of the Euler equations is larger than the Galilei group. In 
this paper we determine, in a systematic manner, the maximal invariance group 
${\cal G}$ of general fluid dynamics and  show that it is a semi-direct product
${\cal G} = SL(2,R) \wedge G$, where the $SL(2,R)$ group contains the 
time-translations, dilations and the inversion $\Sigma$, and $G$ is the static 
(nine-parameter) Galilei group. A subtle aspect of the inclusion of viscosity 
fields is discussed and it is shown that the Navier-Stokes assumption of 
constant viscosity breaks the $SL(2, R)$ group to a two-parameter group of time
translations and dilations in a tensorial way. The 12-parameter group 
${\cal G}$ is also known to be the maximal invariance group of the free 
Schr\"odinger equation. It originates in the free Hamilton-Jacobi equation 
which is central to both fluid dynamics and the Schr\"odinger equation.
\vfill
\hfill DIAS-STP-00-16\hfil\break
\centerline {\bf Introduction}
\bigskip
Considerable efforts are being devoted at present in centres such as the 
National Ignition Facility at the Lawrence Livermore Laboratory, U.S.A., and 
the Laser MegaJoule in Bordeaux, France, to simulate astrophysical systems in 
the laboratory. The motivation for this programme comes from observational 
evidence [1] that the dynamics of the mixing of gases during a supernova 
explosion [2] is very similar to the turbulent splashing and mixing of a 
plasma as a fusion capsule is bombarded by high intensity laser beams [3].  
Since the former physical situation deals with an explosion while the latter 
concerns an implosion, and because the time and length scales involved in the 
two cases are drastically different, it was a challenging task to produce a 
theoretical explanation for the observed similarity.    

In a recent paper [4], Drury and Mendon\c ca produced such an explanation by 
showing that the Euler equations of fluid dynamics which govern both the 
systems are left invariant by an inversion transformation $\Sigma: t\rightarrow
-1/t,~{\bf x}\rightarrow{\bf x}/t$. The discovery of this transformation was 
inspired by similar transformations of the form ${\bf x}\rightarrow f(t)
{\bf x}$ that are used by cosmologists to factor out an arbitrary uniform 
expansion or contraction of the system [5]. As is well-known, the Euler 
equations of fluid dynamics are invariant under the Galilei group [6]. The 
results of [4] show that the maximal kinematical\footnote{${}^1$}{Meaning 
transformations involving only space and time coordinates as opposed to those 
involving internal or gauge degrees of freedom.} invariance group of fluid 
dynamics ${\cal G}$ is larger than the Galilei group [7]. In this paper we 
determine, in a systematic manner, the structure of ${\cal G}$ for general 
fluid dynamics. We also discuss subtleties associated with the inclusion of 
viscosity fields and show that the Navier-Stokes assumption of constant
viscosity leads to a reduction of ${\cal G}$ in a tensorial manner. 

It will be shown that ${\cal G}$ is of the form 
$${\cal G} = SL(2,R)\wedge G\eqno(1)$$
where $G$ is the connected, static Galilei group which induces the 
transformations 
$${\bf x}\quad\rightarrow\quad R{\bf x+ a+ v}t,\qquad~~~ t\rightarrow t
\eqno(2)$$
and $SL(2,R)$ is the group 
$$t\quad\rightarrow\quad {\alpha t + \beta\over \gamma t + \delta},~~~~~~~~
{\bf x}\quad\rightarrow\quad {{\bf x}\over \gamma t + \delta}; ~~~\qquad\alpha
\delta -\beta\gamma = 1\eqno(3)$$
which includes parity, time translations, dilations and the inversion $\Sigma$.

${\cal G}$ is isomorphic to the Niederer group, which is the maximal invariance
group of the free Schr\"odinger equation [8]. The reason for this isomorphism 
is that the kinetic part of the fluid-dynamic Lagrangian (see (17)) is of the 
free Hamilton-Jacobi form, which is the classical limit of the Schr\"odinger 
operator.  
\bigskip
\centerline {\bf The Strategy of Proof}
\bigskip
The general fluid dynamic equations in $n$-dimensional space are [9] 
$$ D\rho = -\rho\nabla\cdot {\bf u} \eqno(4a)$$
$$ \rho D{\bf u} = -\nabla p  + {\bf V} \eqno(4b)$$
$$ D\epsilon = -(\epsilon + p)\nabla\cdot {\bf u}\eqno(4c)$$
where the convective derivative $D$ and the viscosity terms ${\bf V}$ are 
defined by  
$$D = {\partial\over\partial t} + {\bf u}\cdot\nabla~~~~~{\hbox{and}}~~~~~~
V_i = \nabla_j\Bigl(\eta (\nabla_ju_i + \nabla_i u_j - {2\over n}
\delta_{ij}\nabla_k u_k)\Bigr) + \nabla_i (\zeta\nabla_ku_k) \eqno(5)$$
respectively. In the above equations $\rho$, {\bf u}, $p$ and $\epsilon$ stand 
for the density, the velocity vector field, the pressure, and the energy 
density of the fluid respectively and $\eta ,~\zeta$ are the shear and bulk
viscosity fields. The above differential equations are augmented by an 
algebraic condition called the polytropic equation of state which relates the 
pressure to the energy density thus:
$$p = (\gamma_o - 1)\epsilon~~~\Rightarrow~~~p + \epsilon = \gamma_o\epsilon
\eqno(6)$$ 
$\gamma_o$ being a constant called the polytropic exponent. This equation 
may be used to eliminate $p$ from (4). Further, by making the substitution, 
$$\epsilon = \chi\rho^{\gamma_o} \eqno(7)$$
the original set of equations (4) reduces to 
$$ D\rho = -\rho \nabla\cdot {\bf u}\eqno(8a)$$
$$\rho D{\bf u} ~~=~~ -(\gamma_o-1)\nabla(\chi\rho^{\gamma_o}) + 
{\bf V}\eqno(8b)$$
$$ D\chi = 0   \eqno(8c)$$
To find the maximal invariance group of these equations we first note that,
for any system (such as the above) for which there is no feedback from the 
fields to the space-time transformations, the maximal invariance group 
${\cal G}$ of the general class of configurations is also an invariance group 
for any special sub-class of configurations; though not necessarily a 
maximal one, since a restricted set of configurations could have a larger 
symmetry group. Formally, 
$${\cal C}_s \subset {\cal C}\Rightarrow {\cal G}\subseteq{\cal G}_s\eqno (9)$$ 
where ${\cal C}$ denotes general configurations and ${\cal C}_s$ denotes a 
special class. Thus a general strategy for finding maximal invariance groups is
to first find a (tractable) sub-class of configurations ${\cal C}_s$ for which 
the maximal invariance group ${\cal G}_s$ can be found, and then see what 
conditions are imposed on ${\cal G}_s$ by the general configurations. This is 
the strategy we shall adopt; except that in our case we choose a sub-class for 
which the generalisation to arbitrary configurations imposes no further 
conditions {\it i.e.} ${\cal G}_s \equiv {\cal G}$. The tractable sub-class we 
choose is obtained by making three simplifications, namely, ignoring the 
effects of viscosity by tuning the viscosity fields $\eta$ and $\zeta$ to 
zero, by setting $\chi = 1$, and by letting the velocity vector ${\bf u}$ be 
curl-free. The resulting sub-class is the configuration of inviscid, isentropic
and irrotational flows which are well-known in fluid dynamics.  
\bigskip
\centerline{\bf The Sub-class of Inviscid, Isentropic, and Irrotational Fluids}
\bigskip
The first simplification we consider is to ignore viscosity effects in which 
case (8) reduces to 
$$ D\rho = -\rho \nabla\cdot {\bf u}\eqno(10a)$$
$$\rho D{\bf u} = -(\gamma_o-1)\nabla(\chi\rho^{\gamma_o})\eqno(10b)$$
$$ D\chi = 0   \eqno(10c)$$
Second, we note that it is consistent with the field equations to set 
$\chi = 1$, with ${\bf u}$ not necessarily curl-free, in which case the 
equations in (10) reduce to, 
$$ D\rho = -\rho \nabla\cdot {\bf u}\quad \hbox{and} \quad  
D{\bf u} = -\gamma_o(\gamma_o-1)\rho^{\gamma_o-2}\nabla\rho \eqno(11)$$
The reason for making this simplification is that the system becomes a local 
Lagrangian one. This can be seen by using the standard Clebsch parametrization 
[10] 
$${\bf u} = -\nabla\phi-\nu\nabla\theta \eqno(12)$$
for the vector-field ${\bf u}$, in which case the local Lagrangian density is, 
$${\cal L} = \rho\Bigl[\dot\phi+\nu\dot\theta-{1 \over 
2}(\nabla\phi+\nu\nabla\theta)^2 - \rho^{\gamma_o - 1}\Bigr]\eqno(13)$$
It will be convenient to write this Lagrangian density in the form 
$${\cal L} = \rho\Bigl[\dot\phi -{1\over 2}(\nabla\phi)^2+\nu{\cal D}\theta 
-{\nu^2 \over 2}(\nabla\theta)^2- \rho^{\gamma_o - 1}\Bigr]\quad{\hbox{where}} 
\quad {\cal D} = {\partial \over \partial t} - \nabla\phi\cdot\nabla\eqno(14)$$
is the convective derivative $D$ restricted to the curl-free part of ${\bf u}$. 
Clearly $\nu$ is a Lagrange multiplier, and, by varying it, we obtain 
$$\nu = {{\cal D}\theta \over (\nabla\theta)^2} \quad \hbox{and} \quad 
{\cal L} = \rho\Bigl[\dot\phi-{1 \over 2}(\nabla\phi)^2-{ 1\over 2}{
({\cal D}\theta )^2\over(\nabla\theta)^2}- \rho^{\gamma_o - 1}\Bigr]\eqno(15)$$
Note that the first equation here is equivalent to $D\theta=0$ since 
$$D\theta  = {\cal D}\theta  - (\nu\nabla\theta )\cdot\nabla\theta
= {\cal D}\theta - \nu (\nabla \theta )^2 = (\nabla\theta )^2\Bigl(
{{\cal D}\theta\over (\nabla\theta )^2} - \nu\Bigr)\eqno (16)$$
We now proceed to the third and last simplification mentioned.

The third simplification we make is to consider the curl-free case 
${\bf u} = -\nabla\phi$ or $\nu = \theta =0$. Then the Action 
corresponding to the Lagrangian density (13) reduces to   
$$S = \int d^nxdt~ \rho\Bigl[\dot\phi-{1 \over 2}(\nabla\phi)^2 \Bigr] -  
\rho^{\gamma_o}\eqno(17)$$
Note that the quantity in square brackets is just the Hamilton-Jacobi function
for a free particle. Following our strategy we now seek the maximal invariance 
group of the Action (17). 

The most general transformation involving the fields, as shown in the Appendix,
takes the linear, inhomogeneous form 
$$\xi_i = \xi_i(x,t), \qquad \tau = \tau (x,t), \qquad {\tilde\phi} = s(x,t)
\phi + \lambda (x,t),\qquad{\tilde\rho} = \mu(x,t)\rho + \mu' (x,t)\eqno(18)$$
where $s$ is a constant. However, one sees at once that if $\mu'\not=0$ the 
form invariance of the Action is violated by inhomogeneous terms. Furthermore, 
since 
$${\partial\over \partial t}={\partial \tau\over\partial 
t}{\partial\over \partial \tau}+ {\partial \xi_i\over\partial t}
{\partial\over \partial \xi_i} \quad  \hbox{and } \quad 
{\partial\over \partial x_i}={\partial \tau\over\partial x_i}
{\partial\over \partial \tau}+ {\partial \xi_j\over\partial x_i}
{\partial\over \partial \xi_j} \eqno(19)$$  
we see that we obtain a term $(\partial \tilde\phi/\partial \tau)^2$ unless 
$${\partial\tau\over\partial x_i} = 0 \eqno(20)$$ 
This means that the $\partial\phi/\partial t$ term does not pick up an 
$x$-dependent factor relative to $\partial\phi/\partial\tau$. Finally we note 
that under the transformations (18), the second term in the Hamilton-Jacobi 
part of the Action (17) gives
$${1\over 2}h_{ij}{\partial\phi\over\partial x_i}{\partial\phi\over
\partial x_j}~~\rightarrow~~ {1\over 2} g_{ij} {\partial\phi\over\partial 
\xi_i}{\partial\phi\over\partial\xi_j}~~~{\hbox{where}}~~g_{ij} = {\partial
\xi_i\over\partial x_k}{\partial\xi_j\over\partial x_l}h_{kl}~~\equiv~~
\Omega (x) g'_{ij}(\xi ) \eqno(21)$$ 
where $h_{ij}$ is the flat Euclidean metric. As is well-known, the most 
general transformations that have the above property are the special conformal 
transformations with parameter $b$, for which $\Omega =  (1 + 2b\cdot x + 
b^2x^2)^2$, and the Euclidean transformations for which $\Omega = 1$. It is 
easy to see however that, for the Action in (17) to remain invariant, the 
${\partial\phi /\partial t}$ part of the Hamilton-Jacobi function must 
transform in the same way as the $(\partial\phi /\partial x)^2$ part. But since
(20) forbids $\partial\phi /\partial t$ to pick up an $x$-dependent factor, the
conformal transformations in (21) are not allowed. Thus the possible 
transformations reduce to 
$$\xi_i= M_{ij}(t)x_j + w_i(t),\qquad\tau=\tau(t), \qquad \tilde\phi = s\phi + 
\lambda(x,t) \qquad \tilde\rho = \mu(x,t)\rho \eqno(22)$$
where $s$ is a constant and $M$ is Euclidean {\it i.e.} $M = f(t)R(t)$, where 
$f$ is a scale factor and $R(t)$ is a rotation matrix. In that case 
$$\pmatrix{
{\partial \tau \over \partial t} & {\partial \tau \over \partial x_j}\cr 
{\partial \xi_i \over \partial t} & {\partial \xi_i \over \partial x_j} }
=\pmatrix{\dot\tau  & 0 \cr \dot\xi_i   & M_{ij}} \quad \Rightarrow 
\quad \pmatrix{\dot\tau\neq 0\cr {\hbox {det}} M\neq 0 } \eqno(23)$$ 
and the Action becomes 
$$\eqalign{S &= \int {d^n\xi d\tau \over {\hbox {det}}M^n\dot\tau } \mu^{-1}
\tilde\rho\Bigl[ (\dot\tau {\partial \over \partial \tau} + \dot\xi_i{
\partial\over\partial\xi_i}){(\tilde\phi -\lambda)\over s} - {{\hbox{det}}M^2
\over 2s^2} ({\partial \tilde\phi\over \partial\xi_i} - {\partial\lambda\over
\partial\xi_i})^2 - (\mu^{- 1}\tilde\rho)^{\gamma_o - 1}\Bigr] \cr& 
=\int {d^n\xi d\tau\over s\mu {\hbox {det}}M^n}\tilde\rho\Bigl[({\partial
\over \partial \tau} +{\dot\xi_i{\partial\over \partial\xi_i}\over\dot
\tau })(\tilde\phi  - \lambda) - {{{\hbox{det}}M^2\over 2s\dot\tau }}({
\partial\tilde\phi \over\partial\xi_i} - {\partial\lambda\over\partial
\xi_i})^2 - {s(\mu^{-1})^{\gamma_o-1}\over\dot\tau}{\tilde\rho}^{\gamma_o
- 1}\Bigr]}\eqno(24)$$
Invariance then requires that 
$$ s\mu ({\hbox{det}}M)^n = 1, ~~~({\hbox{det}}M)^2 = s\dot\tau,~~~
{\hbox{and}}~~~(\mu^{-1})^{\gamma_o - 1} = {\dot\tau\over s}\eqno(25)$$
and 
$$({\partial \over \partial \tau} + {\partial\xi_i\over\partial\tau}
{\partial\over\partial\xi_i})(\tilde\phi - \lambda) -{{1 \over 2 }} \Bigl(
{\partial\tilde\phi\over\partial\xi} - {\partial\lambda\over\partial\xi}\Bigr)^2
~~~=~~~ {\partial\tilde\phi\over\partial \tau}-{1 \over 2}({\partial\tilde\phi
\over \partial\xi} )^2 \eqno(26)$$
Equations (25) can be readily solved to get 
$$s = 1\Rightarrow ({\hbox{det}}M)^2 = \dot\tau ,~~~\mu = ({\hbox{det}}M)^{-n},
~~~{\hbox {and}}~~~\gamma_o = 1 + {2\over n}\eqno(27)$$
The result for $\gamma_o$ is in agreement with [4]; in particular in 
three-dimensional space, $n = 3$, and $\gamma_o = {5\over 3}$, as expected. 
Substituting these results in (26), and requiring that it be satisfied for all 
$\partial\tilde\phi/\partial \xi_i$, splits it into 
$${\partial\xi_i\over\partial\tau} = -{\partial\lambda\over\partial\xi}_i\quad 
\hbox{and}\quad{\partial\lambda\over\partial\tau} - {1\over 2}\Bigl({\partial
\lambda\over\partial\xi}\Bigr)^2=0,~~~{\hbox{where}}~~~\dot\tau = 
{\hbox{det}}M^2\eqno(28)$$
Notice that the second equation is just the free Hamilton-Jacobi equation with 
Action $\lambda$. The maximal invariance group can now be obtained by solving 
the above equations for the functions $\xi_i$ and $\lambda$. Thus the invariance
group ${\cal G}$ is determined by the structure of the Hamilton-Jacobi equation
for a free particle. However, we will now show, even before solving these 
equations, that the group of invariance remains the same when the 
simplifications made in this section are relaxed and we return to the case of 
the general fluid configurations.  
\bigskip
\centerline{\bf Proof of Maximality for General Fluid Configurations} 
\bigskip
We first  note that the first and last equations in (28) are exactly the 
conditions for the restricted derivative ${\cal D}$ to transform covariantly 
since 
$$\eqalign{ {\cal D} (x, t, \phi ) &={\partial\over\partial t} - {\partial\phi
\over \partial x_i} {\partial\over\partial x_i} = \Bigl(\dot\tau{\partial
\over \partial \tau} +{\partial \xi_i\over \partial t}{\partial \over \partial
 \xi_i} \Bigr) - \Bigl({\partial\xi_j\over\partial x_i}{\partial \xi_k\over
\partial x_i} \Bigr) {\partial\phi\over\partial\xi_j}{\partial\over\partial
\xi_k} \cr& =\Bigl(\dot\tau{\partial\over \partial \tau} - {\hbox{det}}M^2
{\partial\tilde \phi \over \partial\xi_i}{\partial\over\partial\xi_i}\Bigr) +
\Bigl({\partial \xi_i \over \partial t} + {\hbox{det}}M^2{\partial\lambda 
\over \partial \xi_i}\Bigr) {\partial\over\partial\xi_i} = {\hbox{det}}M^2
{\cal D}(\xi ,\tau ,\tilde\phi )}\eqno(29)$$
Let us now relax the assumption that ${\bf u}$ is curl-free, with $\chi =1$
still being true. From the Lagrangian density for this case, 
$$ {\cal L} = \rho\Bigl[\dot\phi-{1 \over 2}(\nabla\phi)^2-{1\over 2}
{({\cal D}\theta )^2\over(\nabla\theta)^2}-\rho^{\gamma_o - 1}\Bigr]\eqno(30)$$
and the covariant transformations, 
$$ {\cal D} (x, t, \phi ) = ({\hbox{det}}M)^2{\cal D}(\xi, \tau , \tilde\phi )
~~~{\hbox{and}} \qquad \nabla (x) = ({\hbox{det}}M)\nabla (\xi )\eqno(31)$$
we see that (30) will be invariant under the given space-time transformations 
provided only that $\nu$ and $\theta$ are scalars. Thus the case of general 
${\bf u}$ puts no restrictions on the space-time transformations but determines
the transformation properties of the $\nu$ and $\theta$ components of 
${\bf u}$. In particular we see that the full convective derivative $D$ and 
${\bf u}$ have the transformation properties  
$$  D (x, t, \phi ) = ({\hbox{det}}M)^2 D(\xi, \tau ,\tilde\phi ),~~~{\hbox
{and}}\qquad \tilde{\bf u} =  {1\over {\hbox{det}}M}\Bigl({\bf u} - \nabla
\lambda \Bigr) \eqno(32)  $$
Note that the inhomogeneous part of the ${\bf u}$-transformation comes from the
curl-free part of ${\bf u}$. Furthermore we see that the ${\cal G}$ 
transformations do not preserve the condition of incompressibility ($\nabla
\cdot{\bf u} = 0$) and hence we need to consider general, compressible fluid
configurations.  

All this is within the Lagrangian framework. In order to relax the $\chi = 1$ 
simplification, we need to allow for general fields $\chi$. For this we must go
outside the Lagrangian framework and consider the field equations. But that is 
simple since the field equations with $\chi = 1$ are invariant and we see by 
inspection that the equations for general $\chi$ remain invariant provided that
$\chi$ transforms as a scalar. Thus the set of equations in (10) is invariant 
with respect to the maximal invariance group provided $\nu ,~\theta$ and 
$\chi$ transform as scalars. Thus the maximal invariance group for $\nu = 
\theta = 0$, $\chi = 1$ (corresponding to our sub-class without viscosity)
is the maximal invariance group for all inviscid configurations. 

Finally we include viscosity effects. It is straightforward to verify that 
the viscosity terms in (4b) are covariant only if $\nabla\cdot{\bf u} = 0$
or $\nabla_i\zeta = 0$. However, as already mentioned, the former condition 
is not invariant under ${\cal G}$ transformations and hence covariance 
requires that $\zeta$ is a constant in space (although not in time). In this 
case, the fluid equations (4) would remain invariant if the viscosity fields 
transform as tensors of rank $n$, where $n$ is the space dimension {\it i.e.} 
$\eta\rightarrow {\hbox{det}}M^n \eta$ and  $\zeta\rightarrow {\hbox{det}}M^n
\zeta$. This in turn implies that the equations would remain invariant under 
the Navier-Stokes assumption of constant viscosity only if det$M$ is a constant
which breaks the ${\cal G}$ symmetry in a tensorial way. Therefore the maximal
invariance group of the Navier-Stokes equations contains just the static 
Galilei group plus dilations and time translations corresponding to det$M$ 
being a constant, as will be clear when we solve for the ${\cal G}$ 
transformation functions in the next section. 

It is worth mentioning that the breakdown of ${\cal G}$ by the Navier-Stokes
restriction does not contradict (9) because, in contrast to the other 
restrictions $\eta=\zeta=0$, $\chi=1$, ${\bf\nabla\times u}=0$ that we have 
considered, this restriction has a feed-back effect on the non-linear part of 
the space-time transformations.  
\bigskip
\centerline{\bf The Solutions for the Transformation Functions} 
\bigskip
Returning to the equations in (28), it is first useful to note from the 
definition of $\xi$ in (22) that  
$${\partial\xi_i\over \partial\tau} = A_{ij}\xi_j + W_i~~{\hbox{ where}}~~ 
A\equiv M_\tau M^T =  f_\tau f^{-1} +  R_\tau R^T~~ {\hbox{and}}~~ W  = M{
\partial\over\partial\tau}(M^{-1}w) \eqno(33)  $$
where the subscript $\tau$ denotes differentiation with respect to $\tau$.
Hence the first equation in (28) becomes 
$${\partial\lambda\over \partial\xi_i} = -A_{ij}\xi_j - W_i~~~~
 \Rightarrow ~~~~\Bigl({\partial^2\lambda\over\partial\xi_i
\partial \xi_j} = -A_{ij}\Bigr)\eqno(34)$$ 
Here the equation in the brackets shows that $A$ is a symmetric matrix. But 
since $R$ is a rotation matrix, $R_\tau R^T$ is in the Lie algebra of the 
rotation group and hence is antisymmetric. Thus we have the result $R_\tau = 0$
and hence $R$ is a constant (rigid) rotation matrix.\footnote{$^2$}{With this 
in mind, we suppress the matrix for convenience, and restore it later.} 

\noindent It then follows that det$M = f$ and from (28) and (32) we have 
$$\dot\tau = f^2~~~{\hbox{and}}~~~\tilde{\bf u} = {1\over f}\Bigl({\bf u} - 
\nabla\lambda\Bigr) \eqno(35)$$
 The definitions of $A$ and $W$ then simplify to
$$A = f_\tau f^{-1}~~~{\hbox{and}}~~~ W_i =  f\partial_\tau\Bigl({w_i
\over f}\Bigr)\eqno(36)$$
Integrating (34) with respect to $\xi$ we have  
$$\lambda = -A{\xi^2\over 2} - W_i\xi_i - h (\tau ) \eqno(37)$$
Substituting this into the second equation in (28) we get 
$$-{(\xi )^2\over 2}{\partial A\over\partial\tau} - \xi_i{\partial W_i\over
\partial\tau} - {\partial h\over\partial\tau}~~ = ~~{1\over 2}(A\xi + W)^2
\eqno(38)$$
Comparing the coefficients of the powers of $\xi$ breaks this into 
$${\partial A\over\partial \tau} = -A^2, \qquad {\partial W_i\over\partial\tau} 
= - AW_i = -{\partial_\tau f \over f}W_i, \qquad {\partial h\over\partial \tau}
= -{W^2\over 2} \eqno(39)$$
We can solve the second equation in (39) explicitly to get 
$$W_i= {v_i\over f}~~~\Rightarrow~~~ \partial_t \Bigl({w_i\over f}\Bigr) = v_i 
~~~\Rightarrow~~~ w_i = f(t)(v_it + a_i) \eqno(40)$$
where the $v_i$ and $a_i$ are constants. It follows that 
$$\xi_i~\equiv~fx_i + w_i~= ~f(t) (x_i + a_i + v_it)\eqno(41)$$ 
The third equation in (39) can then be written as 
$${\partial h\over\partial \tau} = -{v^2\over 2f^2}
~~\Rightarrow~~ {\partial h\over\partial t} = -{v^2\over 2}
~~\Rightarrow~~ h = h_0 - {v^2\over 2}t\eqno(42)$$
Since the transformations of the fields are given by $\tilde\rho = f^{-n}\rho$
and $\tilde\phi = \phi + \lambda$, we see that everything is determined by the
function $f$ which satisfies 
$${\partial\over\partial \tau}\Bigl({\dot f \over f^3}\Bigr)
=-\Bigl({\dot f\over f^3}\Bigr)^2 \eqno(43)$$
Using $\dot\tau = f^2$, this is easily seen to be equivalent to 
$$\partial_\tau^2 f = 0~~~{\hbox{and}}~~~{\dddot\tau\over\dot\tau} - {3\over 2}
\Bigl({\ddot\tau\over\dot\tau} \Bigr)^2 = 0 \eqno(44)$$ 
The left hand side of the second equation is the Schwarzian derivative of 
$\tau$ and it follows that the general solution of the above equation is 
$$\tau = {\alpha t + \beta\over \gamma t + \delta}~~~{\hbox{and}}~~~
f = {1\over \gamma t + \delta}~~~{\hbox{where}}~~~\alpha\delta - \beta\gamma
= 1\eqno(45)$$
Thus the most general transformations of the time coordinate form an $SL(2, R)$
group. The second equation in (35) then reduces to 
$$\tilde {\bf u} = (\gamma t + \delta){\bf u} - \gamma ({\bf x + a}) + 
\delta{\bf v} \eqno(46)$$
From (41) it also follows that 
$$\xi_i~~= ~~{1\over\gamma t + \delta}(x_i + a_i + v_it) \eqno(47)$$ 

Returning to the general case, and using the inversions 
$$\gamma\tau - \alpha = {-1\over\gamma t + \delta}\quad\hbox{and}\quad  
\gamma\xi - v = {\gamma (x+a) - \delta v\over\gamma t +\delta}\eqno(48)$$
we have 
$$\lambda = \lambda_0 - {(\gamma\xi - v)^2 \over 2\gamma(\gamma\tau - \alpha )} 
= \lambda_0 + {[\gamma(x + a) -\delta v]^2 \over 2\gamma(\gamma t + \delta)} 
\quad\hbox{where}\quad\lambda_0 = -\Bigl(h_0 + {\delta\over\gamma}{v^2\over 2}
\Bigr)\eqno(49)$$
The case ${\hbox{det}}M = f =$ a constant corresponds to letting $\gamma = 0$
and from (45) and (47) it is clear that the group of transformations in this 
case includes the static Galilei group $G$, the dilations, the time 
translations and parity, but excludes time-reversal. As already mentioned, this
is the maximal invariance group of the Navier-Stokes equations. It is also 
useful to consider the following two special cases. 

\noindent {\it I. $\beta = \gamma = 0,~\alpha = 1$: The Connected, Static 
Galilei Transformations:}
In this case, we have 
$$g:~~~\tau = t,~~~~~{\bf \xi} = R{\bf x + a + v}t \eqno(50)$$
where we have restored the rotations, and from (49) it follows  
$$\lambda = -h_0 - {\bf v} \cdot (R{\bf x + a}) - {v^2\over 2}t\eqno(51) $$
These equations describe connected, static Galilei transformations which 
exclude parity and time-reversal. 

\noindent{\it II. ${\bf a = v = 0},~ R = 1$: The Inversion Transformations:} 
In this case, we have 
$$\sigma:~~~~~\tau = {\alpha t + \beta\over\gamma t + \delta},\qquad {\bf\xi} = 
{R{\bf x}\over\gamma t + \delta};~~~~\alpha\delta - \beta\gamma = 1\eqno(52) $$
These are the $SL(2,R)$ generalisations of the inversion transformations 
presented in [4]. For the transformations of the fields we have from (49) 
$$\lambda =  {\gamma x^2\over 2(\gamma t +\delta)} - h_0\eqno (53)$$
\bigskip
\centerline{\bf The Maximal Invariance Group ${\cal G}$}
\bigskip
To understand the structure of the group, we study the relationship between the
$SL(2, R)$ group and the connected static Galilei group $G$. Let us first 
consider a conjugation of a $g\in G$ by a $\sigma\in SL(2, R)$. By making 
three successive transformations of $x$ and $t$ we find that 
$$\sigma^{-1}(\alpha , \beta , \gamma )g(R, {\bf a}, {\bf v}) \sigma (\alpha ,
\beta , \gamma ) = g(R, {\bf a_\sigma,  v_\sigma} )\eqno(54)$$
where 
$$\pmatrix{{\bf a_\sigma}\cr {\bf v_\sigma} }
= \pmatrix {\delta & \beta\cr
\gamma & \alpha } 
\pmatrix{{\bf a}\cr {\bf v}}\eqno(55)$$
This shows that $G$ is an invariant sub-group and so the group structure is 
$${\cal G} = SL(2, R)\wedge G\eqno(56)$$
where $\wedge$ denotes semi-direct product with $G$ as invariant subgroup. 
More precisely, if we recall that $G$ itself takes the form 
$$ G = R\wedge \Bigl( T({\bf a}) \otimes B({\bf v})\Bigr)\eqno(57)$$
where $T$ and $B$ are the translation and boost groups with parameters 
${\bf a}$ and ${\bf v}$ respectively, then we see that $SL(2, R)$ commutes 
with $R$ and mixes $T$ and $B$ in the manner shown in (55). 

The original inversion $\Sigma$ is the special element of $SL(2, R)$ for 
which $(\alpha ,~ \beta ,~ \gamma ,~ \delta ) = (0,~ -1,~ 1,~ 0)$. Note that 
$\Sigma^2 = P$ where $P$ is the parity. Furthermore if we consider the 
coset elements $g_{\atop\Sigma}(R,~ {\bf a},~ {\bf v}) \equiv \Sigma 
g(R,~{\bf a},~{\bf v})$, where $g \in G$, we have using (54) and the 
standard property of Galilei transformations {\it viz.} 
$g\bigl(R, {\bf a}, {\bf v})g(R', {\bf a}', {\bf v}'\bigr) = 
g\bigl(RR', R{\bf a}' + {\bf a}, R{\bf v}' + {\bf v}\bigr)$,  
$$\eqalign{g_{\atop\Sigma} (R',~ {\bf a}',~ {\bf v}')g_{\atop\Sigma}(R,~{\bf 
a},~{\bf v}) &= g_{\atop P}(R'R,~R'{\bf a -  v}',~R'{\bf v + a}')\cr
\Rightarrow~~~ g_{\atop \Sigma}^2(R,~{\bf a,~v}) &= g_{\atop P}(R^2, R{\bf a -
 v}, R{\bf v +  a})}\eqno(58)$$
where we have used the obvious notation $g_{\atop P} (R, ~{\bf a, ~v}) = Pg
(R,~{\bf a, v})$. It follows from the above equation that 
$$g_{\atop\Sigma}^4(R,~{\bf a,~v}) = g\Bigl(R^4,~(R^2-1)(R{\bf a -v}),~(R^2-1)
(R{\bf v +a})\Bigr) \eqno(59)$$
Since $R{\bf a - v}$ and $R{\bf v+a}$ are linearly independent, this shows that
every connected Galilei transformation is the fourth power of a coset 
transformation.  
\bigskip
\vfil\eject
\centerline{\bf Extension to the Quantum Theory} 
\bigskip
If a classical system is described by the Galilei group $G$ in (57) then, in 
the field theoretic representation, there is a central extension of $G$; 
namely, the well-known one-parameter mass group whose generator commutes with 
all the generators of $G$. Since a quantum wavefunction can be thought of as a 
nonrelativistic field, the corresponding quantum system is described by a 
central extension of the group of the form  
$$G = R\wedge \Bigl(T ({\bf a}), B({\bf v})\Bigr)\eqno(60)$$
where the group $(T, B)$ is no longer abelian, but is a Heisenberg-Weyl group 
of the form 
$$ T({\bf a})B({\bf v}) = B({\bf v})T({\bf a})e^{i\hbar M{\bf  a}\cdot{\bf v}}
\eqno(61)$$ 
and $M$ is a central constant. It is obvious that this relation is invariant 
with respect to the transformation (55) of the parameters ${\bf a}$ and 
${\bf v}$ induced by $SL(2, R)$. Thus ${\cal G}$ remains a symmetry group of 
the quantised system.  
\bigskip
\centerline{\bf Connection to the Free Schr\"odinger Equation} 
\bigskip
The fact that ${\cal G}$ is a covariance group of the Hamilton-Jacobi function 
$\partial\phi/\partial t + 1/2(\nabla\phi)^2$ shows that it is an invariance 
group of a free particle. The above discussion of the quantum extension implies
that it is also an invariance group of a quantised free particle. Indeed it was
shown in 1972 that ${\cal G}$ is the maximal invariance group of the free 
particle Schr\" odinger equation [8]. Since the invariance under the Galilei 
group is well-known, it suffices to verify that the Schr\"odinger equation    
$$i\hbar {\partial\psi\over \partial t} + {\hbar^2\over 2m}{\partial^2\psi
\over\partial x_i^2} = 0\eqno(62)$$ 
remains form-invariant under the ${\cal G}$. It is easily checked that this
is accomplished by the following transformation of the wavefunction
$$\psi ({\bf x}, t)~~ \sim~~ (\gamma\tau - \alpha )^{n\over 2} e^{-i\lambda}
\psi({\bf \xi} , \tau)~~\sim~~ (\gamma t + \delta)^{-{n\over 2}}e^{-i\lambda}
\psi\eqno(63)$$
where $\lambda$ is defined in (49). Note that at time $\tau = \infty$, {\it 
i.e.} $t = -{\delta\over\gamma}$, the wavefunction becomes infinite, but this 
is precisely the singularity associated with the explosion. 
\bigskip
\centerline{\bf Conclusions} 
\bigskip
We have investigated the maximal kinematical invariance group of general fluid 
mechanics and have found that it is a semi-direct product of the form 
$SL(2,R)\wedge G$ where $G$ is the static Galilei group. The incompressibility
condition is not preserved by the above transformations and hence the 
viscosity terms transform covariantly only if the viscosity field $\zeta$ is 
constant in space (although not in time). In this case the fluid dynamic 
equations remain invariant if the viscosity fields transform like tensors of 
rank $n$, where $n$ is the space dimension. The Navier-Stokes assumption of 
constant viscosity breaks the $SL(2, R)$ part of ${\cal G}$ to a two-parameter
group of dilations and time translations in a tensorial manner.

The inversion transformation $\Sigma$ found in [4] is a special element of the 
$SL(2, R)$ and acts like the square root of parity. The transformations 
generated by the coset elements $g_{\atop \Sigma} = \Sigma g$ act like the 
fourth roots of connected Galilei transformations. It is also pointed out that 
${\cal G}$ is the Niederer group which is the maximal invariance group of the 
free Schr\"odinger equation. The reason for this is that ${\cal G}$ is actually
the invariance group of the free Hamilton-Jacobi function, which is common to 
both fluid dynamics and the classical limit of the Schr\"odinger equation. It 
is surprising that such a basic result does not seem to be more widely known! 
\bigskip
\centerline{\bf Appendix} 
In this appendix we show that the linear inhomogeneous transformations of the 
field variables considered in (18) are indeed the most general field 
transformations allowed. In order to show this we begin by noting that the 
most general transformations can be written as follows:
$$\tilde\phi \equiv \tilde\phi (\xi, \tau, \phi )~~~{\hbox{and}}~~~
\tilde\rho\equiv \tilde\rho (\xi, \tau, \rho ) \eqno(A1)$$ 
where $\tilde\phi$ and $\tilde\rho$ are {\it a priori} arbitrary functions 
of the old fields $\phi$ and $\rho$. Note that we are not allowing the fields 
to mix, but this is reasonable since $\phi$ and $\rho$ are really not on the 
same footing; whereas $\phi$ appears in the Action (17) only through its
derivatives, $\rho$ appears without any derivatives. Using the fact that
$x,~t\rightarrow \xi, \tau$, the Hamilton-Jacobi function can be rewritten as
follows:
$${\partial\phi\over\partial t} - {1\over 2}\bigl({\partial\phi\over\partial x}
\bigr)^2~~ =~~ \bigl({\partial\xi\over\partial t}{\partial\phi\over\partial\xi}
 + {\partial\tau\over\partial t}{\partial\phi\over\partial\tau}\bigr) - 
{1\over 2}\bigl({\partial\xi\over\partial x}{\partial\phi\over\partial\xi} + 
{\partial\tau\over\partial x}{\partial\phi\over\partial\tau}\bigr)^2\eqno(A2)$$ 
Now it follows from  
$$\tilde\phi~\equiv~\tilde\phi (\xi, \tau, \phi )~~ \Rightarrow~~ \phi~ \equiv
~F(\xi, \tau, \tilde\phi )\eqno(A3)$$  
that
$${\partial\phi\over\partial\xi} = {\partial F\over \partial\xi} + 
{\partial\tilde\phi\over\partial\xi}{\partial F\over\partial\tilde\phi}
~~~~{\hbox{and}}~~~~ 
{\partial\phi\over\partial\tau} = {\partial F\over \partial\tau} + 
{\partial\tilde\phi\over\partial\tau}{\partial F\over\partial\tilde\phi}
\eqno(A4)$$
We would now like to fix the form of $F$ by requiring the covariance of the 
Hamilton-Jacobi function {\it i.e.} 
$${\partial\phi\over\partial t} - {1\over 2}\bigl({\partial\phi\over\partial x}
\bigr)^2~~\propto~~{\partial\tilde\phi\over\partial\tau} - {1\over 2}\bigl({
\partial\tilde\phi\over\partial\xi}\bigr)^2$$
Substituting $(A4)$ on the right hand side of $(A2)$, we notice that 
we get terms of the form $(\partial\tilde\phi /\partial\tau )^2$ unless 
$\partial\tau /\partial x = 0 \Rightarrow \tau \equiv \tau (t)$. Equating the 
coefficients of $\partial\tilde\phi/\partial\tau$ and $-(1/2)(\partial\tilde
\phi /\partial\xi)^2$ because of covariance, we get  
$${\partial F \over\partial\tilde\phi} = {{\partial\tau\over\partial t}\over 
\bigl({\partial\xi\over\partial x}\bigr)^2}\eqno(A5)$$
Requiring the coefficient of the $\partial\tilde\phi /\partial\xi$ term to 
vanish produces  
$$ {\partial F\over\partial\xi} = {{\partial\xi\over\partial t}\over  
\bigl({\partial\xi\over\partial x}\bigr)^2}\eqno(A6)$$
Similarly requiring the coefficient of the $\tilde\phi$--independent terms to 
vanish we get 
$${\partial\xi\over\partial t} {\partial F\over\partial\xi} + {\partial\tau
\over\partial t} {\partial F\over\partial\tau} - {1\over 2} \bigl({\partial
\xi\over\partial x}\bigr)^2\bigl({\partial F\over\partial\xi}\bigr)^2$$
Substituting this equation in $(A6)$ gives, after some algebra, 
$${\partial F\over\partial\tau} = -{1\over 2}{\bigl({\partial\xi\over\partial 
t}\bigr)^2 \over\bigl({\partial\xi\over\partial x}\bigr)^2
{\partial\tau\over\partial t}}\eqno(A7)$$
From $(A5)$ it follows that since $\tilde\phi$ is arbitrary, the left hand 
side changes whereas the right hand side, for a given $\xi , \tau , x, t$, 
is a constant in the field. Hence
$${\partial F\over\partial\tilde\phi } \equiv s^{-1}(\xi, \tau )\eqno(A8)$$
where $s^{-1}$ is a constant in the field variable. Solving the above 
differential equation we get 
$$ F = s^{-1}\tilde\phi + \tilde\lambda (\xi, \tau ) \Rightarrow 
\tilde\phi = s(\xi, \tau )\phi + \lambda(\xi , \tau )\eqno(A9)$$
where $\lambda \equiv -s\tilde\lambda$. Further by taking a derivative of 
$(A9)$ with respect to $\xi$ we see that 
$${\partial F\over\partial\xi} = {\partial s^{-1}\over\partial \xi}\tilde\phi
+ s^{-1}{\partial\tilde\phi\over\partial\xi} + {\partial\tilde\lambda\over
\partial \xi}\eqno(A10)$$ 
Again from $(A6)$ it is clear that the left hand side of the above equation 
is independent of $\tilde\phi$ and hence, $s^{-1}$ is independent of $\xi$.
Similarly, by appealing to $(A7)$ we can show that it is independent of $\tau$.
Hence $s$ is a constant. A similar analysis may be carried out to establish 
the generality of the $\rho$ transformations in (18). 
\bigskip
\bigskip
\centerline{\bf Acknowledgements} 
\bigskip
We thank Luke Drury for introducing us to this topic and for many useful 
discussions, David Saakian for raising the question of viscosity, O. Jahn
for his interest, and R. Jackiw for valuable correspondence. 
\vfil\eject
\centerline {\bf References}
\bigskip
\item {1. } Supernova Hydrodynamics Up Close: Science and Technology Review,
Jan'/Feb' 2000; http://www.llnl.gov/str
\item {2. } I. Hachisu {\it et al}, Astrophysical Journal, {\bf 368}, (1991), 
L27.
\item {3. } H. Sakagami and K. Nishihara, Physics of Fluids {\bf B 2}, (1990), 
2715.
\item {4. } L. O'C Drury and J. T. Mendon\c ca, Physics of Plasmas {\bf 7}, 
(2000) 5148. 
\item {5.} S. Weinberg, Gravitation and Cosmology; John Wiley and Sons Inc.
(1972).  
\item {6. } E. C. G. Sudarshan and N. Mukunda, Classical Dynamics: A Modern 
Perspective, John Wiley \& Sons (1974). 
\item {7. } For related work in this matter see, 
M. Hassaine and P. A. Horvathy, Ann. of Phys. {\bf 282} (2000) 218;
Phys. Lett {\bf A279} (2001) 215; A. M. Grundland and L. Lalague, Can. J. 
Phys. {\bf 72}, (1994) 362, Can. J. Phys. {\bf 73}, (1995) 463; R. Jackiw, 
physics/0010042.
\item {8. } U. Niederer, Helvetica Physica Acta, {\bf 45} (1972) 802; 
C. R. Hagen Phys. Rev. {\bf D5} (1972) 377; R. Jackiw, Phys. Today {\bf 25} 
(1972) 23. 
\item {9. } L. D. Landau and E. M. Lifshitz, Fluid Mechanics, Pergamon Press
(1959).
\item {10. } A. Clebsch, Journal f\" ur die reine und angewandte Mathematik,
{\bf 56} (1859), 1; H. Lamb, Hydrodynamics, Cambridge University Press, 1942.
{\it For related work, see} S. Deser, R. Jackiw and A. P. Polychronakos,
physics/0006056; R. Jackiw, V. P. Nair and So-Young Pi, hep-th/0004084. 
\end